\documentclass[%
reprint,
superscriptaddress,
amsmath,amssymb,
aps,
longbibliography,
]{revtex4-1}
\usepackage{amsmath}
\usepackage{amssymb}
\usepackage{graphicx}
\usepackage{braket}
\usepackage{color}
\usepackage{xcolor}
\usepackage{siunitx}
\usepackage{upgreek}
\usepackage[utf8]{inputenc}
\usepackage{physics}
\usepackage{multirow}
\usepackage{array}
\usepackage{url}
\usepackage{dcolumn}
\usepackage{bm}
\usepackage{hyperref}
\usepackage{natbib}

\begin{document}



\title{Symmetry and Control of Spin-Scattering Processes in \\Two-Dimensional Transition Metal Dichalcogenides}

\author{Carmem~M.~Gilardoni}
\email{c.maia.gilardoni@rug.nl}
\author{Freddie~Hendriks}
\author{Caspar~H.~van~der~Wal}
\author{Marcos~H.~D.~Guimar\~aes}
\email{m.h.guimaraes@rug.nl}
\affiliation{Zernike Institute for Advanced Materials, University of Groningen, NL-9747AG  Groningen, The Netherlands}

\date{Version of \today}


\begin{abstract}
Transition metal dichalcogenides (TMDs) combine interesting optical and spintronic properties in an atomically-thin material, where the light polarization can be used to control the spin and valley degrees-of-freedom for the development of novel opto-spintronic devices.
These promising properties emerge due to their large spin-orbit coupling in combination with their crystal symmetries.
Here, we provide simple symmetry arguments in a group-theory approach to unveil the symmetry-allowed spin scattering mechanisms, and indicate how one can use these concepts towards an external control of the spin lifetime.
We perform this analysis for both monolayer (inversion asymmetric) and bilayer (inversion symmetric) crystals, indicating the different mechanisms that play a role in these systems.
We show that, in monolayer TMDs, electrons and holes transform fundamentally differently – leading to distinct spin-scattering processes.
We find that one of the electronic states in the conduction band is partially protected by time-reversal symmetry, indicating a longer spin lifetime for that state.
In bilayer and bulk TMDs, a hidden spin-polarization can exist within each layer despite the presence of global inversion symmetry. We show that this feature enables control of the interlayer spin-flipping scattering processes via an out-of-plane electric field, providing a mechanism for electrical control of the spin lifetime.
\end{abstract}

\maketitle


\textbf{1.} Thin layers of transition metal dichalcogenides (TMD) offer the possibility of electrically and optically addressing spin, valley and layer degrees of freedom of charge carriers \cite{Xiao2012, Mak2012, Zeng2012, Xu2014, Wang2018}. This has led to increased interest in these materials for applications in novel electronic and spintronic devices \cite{Mak2016, Liu2016, Schaibley2016, Zhong2017, Luo2017, Avsar2017}. These properties arise from the symmetries of these intrinsically two-dimensional crystals, combined with the large spin-orbit coupling imprinted on electrons by the heavy transition metal atoms in the lattice \cite{Xiao2012, Xu2014, Ribeiro-Soares2014}.
Several experimental and theoretical works explore the spin and valley lifetimes in monolayer, bilayer, and bulk TMDs, with often contrasting results. In the particular case of spin lifetime in TMDs, experimental values span over 5 orders of magnitude \cite{Lagarde2014, Mai2014, Hsu2015, Yang2015, Ersfeld2019, Song2016}.
Group-theory-based analysis of the symmetries in this class of materials has been useful in unraveling their optical \cite{Lagarde2014, Robert2017, Robert2020} and spintronic \cite{Song2013, Kormanyos2018, Forste2020} properties, including how electrons, holes and excitons couple to phonons and external magnetic fields, for example.
This approach can be very powerful to connect and compare seemingly contrasting results, as well as giving powerful symmetry-based predictions for the design of future experiments.
However, literature still lacks a pedagogical derivation of the mechanisms leading to spin scattering in monolayer and, particularly,  bilayer TMDs based solely on the symmetry of these materials.
Moreover, a careful analysis and understanding of the impact of crystal symmetries on the spintronic properties of these materials can lead to better device engineering which exploit symmetry breaking for active control over the spin information.

\textbf{2.} Here, we apply group-theoretical considerations to obtain the symmetry of the electronic wavefunctions at the edges of the bands in these semiconductors, for both monolayer and bilayer systems. 
In order to do this, we use double groups to unravel the transformation properties of the Bloch wavefunctions including spin at the high symmetry points in the Brillouin zone (BZ), in the absence of external fields. 
Based on these results, we derive the first-order selection rules for spin-scattering processes in a single-particle picture. 
This allows us to determine how electron and hole spins couple to phonons and external fields, and which mechanisms dominate spin-flipping processes at low temperatures. Based on these results, we find that electrons and holes in these materials transform differently. 
In particular, a combination of rotational symmetry and strong spin-orbit coupling (SOC) strongly suppresses low-temperature spin scattering mechanisms for conduction-band electrons in monolayer TMDs. 
For bilayer (and few-layer) systems where individual layers are partially decoupled, we find that an electric field enhances interlayer spin-scattering processes, enabling electrical control of an optically created spin polarization. 
Despite being based on several approximations, the group-theoretical framework developed here allows us to intuitively understand various spin properties of this class of materials in a straight-forward manner, and in line with recent experimental results.

\textbf{3.} This paper is organized as follows: in part I, we focus on monolayer TMDs and their symmetries. 
We obtain the transformation properties of the Bloch wavefunctions including spin at the high-symmetry points of the BZ. 
Based on the symmetries of these wavefunctions, we derive which perturbations (electromagnetic fields and lattice phonon modes) can couple eigenstates with opposite spin, which allows us to determine the processes most likely to lead to spin flips at low temperatures. 
In part II, we repeat this analysis for bilayer TMDs. 
Finally we summarize the main conclusions and elaborate on the impact of our findings to past and future experiments in the field.

\section{Monolayer TMD}
\label{sec::ML}

\subsection{Symmetries of the spatial eigenstates}
\textbf{4.} In order to derive the spin-scattering selection rules at the edges of the bands, one must first obtain the symmetry properties of the eigenstates at the K and K' points of the BZ. 
These properties are determined by the point group describing the crystallographic symmetry at these points, the orbital character of the wavefunctions and the spin of the charge carriers in these states \cite{Ribeiro-Soares2014}. 
In this way, we can classify the electronic eigenstates at the band edges in these materials by their transformation properties, which are summarized by the irreducible representation (IR) of the suitable point group.

\begin{figure}[t]
  \includegraphics[width=\columnwidth]{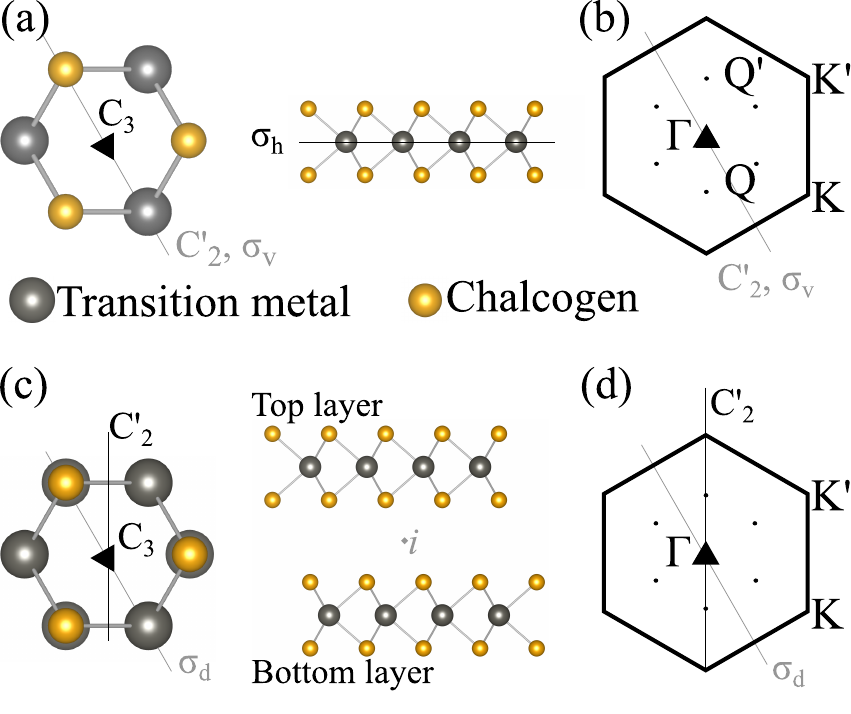}
  \caption{\textbf{Symmetry of the group of the wave vector at the K and K' points}. Lattice structure in real and reciprocal space for monolayer (a-b) and bilayer (c-d) TMDs. The symmetry of the full crystal (that is, the symmetry at the $\Gamma$ point in the BZ) is D\textsubscript{3h} for the monolayer, and D\textsubscript{3d} for the bilayer, for which the symmetry operations are shown explicitly. Operations that, although present in the $\Gamma$ point of the BZ, are absent in the K and K' points (b,d) are shown in light gray. The absence of these symmetry operations at the K and K' points reduces the symmetry at these points to the point groups C\textsubscript{3h} for monolayers and D\textsubscript{3} for bilayers.
  \label{Fig::geom}}
\end{figure}

\textbf{5.} A TMD monolayer (ML) is composed of transition metal and chalcogen atoms, arranged in a hexagonal lattice. Although the coordination between these atoms can vary, the most widely studied TMD polytypes (2H types) have a transition metal atom bound to 6 chalcogen atoms in a trigonal prismatic geometry (Fig.~\ref{Fig::geom}a), giving rise to the crystallographic point-group D\textsubscript{3h}. 
However, the edges of the valence and conduction bands in ML TMDs are located at the K and K' points of the BZ, where not all symmetries of the lattice are preserved. 
Here, only the three-fold rotational symmetry axis (C\textsubscript{3}), the horizontal mirror plane ($\upsigma _{\text{h}}$), and their combinations are valid symmetry operations, such that the wavefunctions at the K and K' points of the BZ transform according to the point-group C\textsubscript{3h} \cite{Song2013, Ribeiro-Soares2014, Robert2017}. 
Figure \ref{Fig::geom}a shows the symmetry operations at the K and K' points in black, and the additional symmetry operations in the $\Gamma$ point in gray.

\begin{figure*}[t]
  \includegraphics[width=\textwidth]{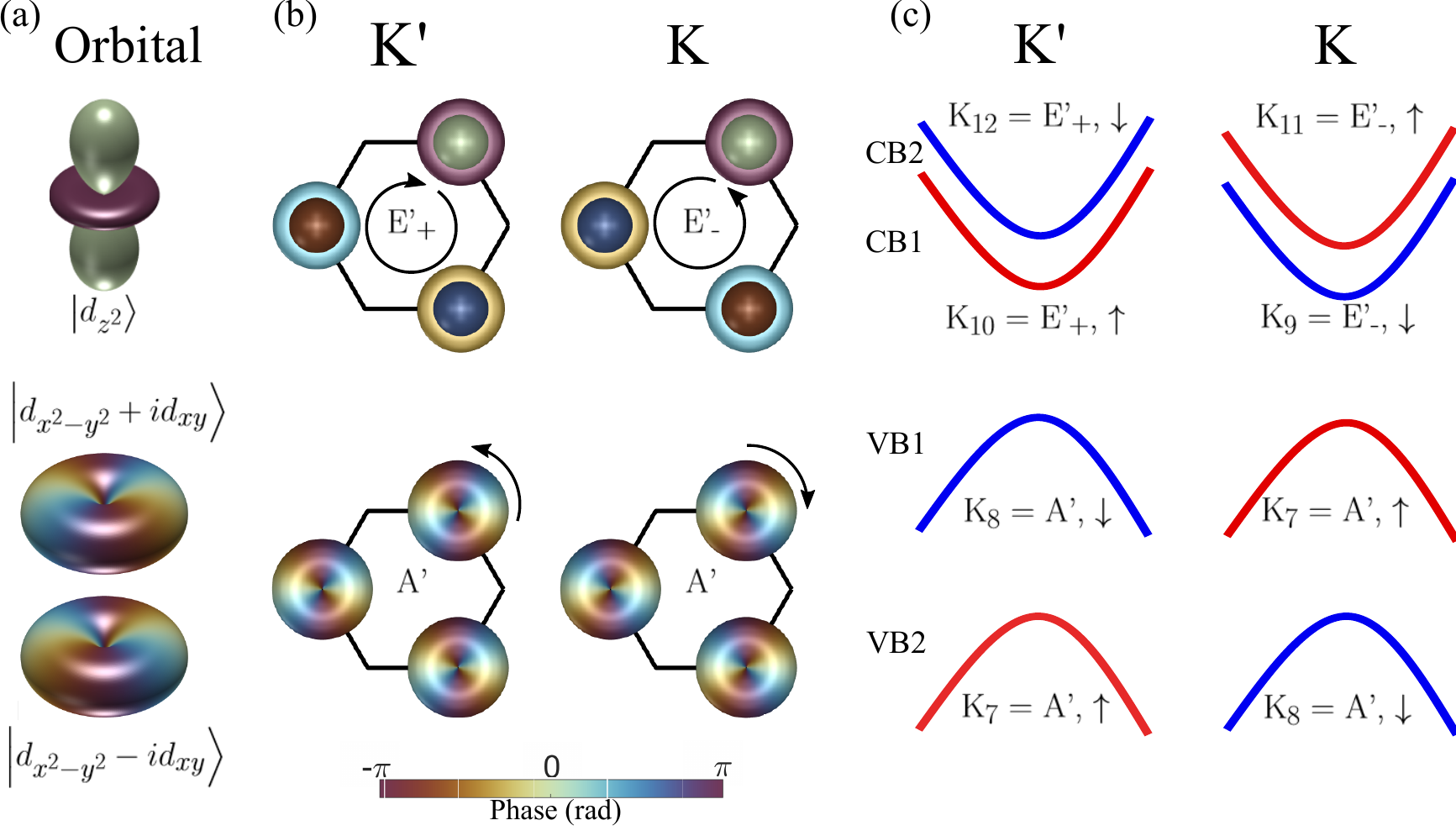}
  \caption{\textbf{Electronic wavefunctions at the K and K' points.} We can obtain the symmetry of the wavefunctions at the K and K' points by applying the symmetry operations of the system to the atomic d-orbitals localized around the TM atoms (a). In a monolayer, the electronic wavefunction at the edge of the conduction band is mainly composed of TM $\ket{d_{z^2}}$ atomic orbitals, which has constant azimuthal phase. (In (a) and (b), the magnitude of the wavefunction in real space is indicated by the surface, whereas the color corresponds to the azimuthal phase according to the scale in the color bar.) When considering also the phase acquired due to translation, the electronic wavefunction at the K (K') point transforms as the E'\textsubscript{+} (E'\textsubscript{-}) (b, top). In the valence band, the electronic wavefunction at the K and K' points is mostly composed of TM $\ket{d_{x^2-y^2}}$ and $\ket{d_{xy}}$ atomic orbitals, which can be combined into the spherical harmonics with $L = 2, m_l = \pm 2$. The phases acquired due to rotation of the spherical harmonics and translation cancel out, to give a final state that transforms as A' in both K and K' points (b, bottom). When considering also the properties of the spin under rotation, we obtain the symmetry of the spin-orbit split wavefunctions in valence and conduction bands by multiplying the IRs of (b) with the irreducible representations associated with spin up and down (K\textsubscript{7} and K\textsubscript{8}, respectively) (c). Hybridization with other orbitals will not change the particular symmetry of the wavefunctions.
  \label{Fig::wavefunction}} 
\end{figure*}

\textbf{6.} \textit{Ab-initio} calculations and tight-binding models of these materials show that the orbital character of the electronic wavefunctions at the K and K' points are largely composed of the d-orbitals of the transition metal atoms \cite{Silva-Guillen2016, Kormanyos2015,Liu2013}.
The valence band wavefunctions are composed predominantly of linear combinations of $d_{x^2-y^2}$ and $d_{xy}$ orbitals, while the conduction band wavefunctions are composed predominantly of the $d_{z^2}$ orbital localized at the transition metal atoms.
Based on this, we can visualize the transformation properties of the wavefunctions at the edges of the valence and conduction bands (Fig.~\ref{Fig::wavefunction}a).
To obtain the symmetry adapted eigenstates delocalized through the lattice, one takes the wavefunction centered at a single transition metal atomic-site and performs on it all symmetry-group operations. 
Due to the nonzero momentum at the K and K' point, a symmetry operation that changes the atomic-site of the orbital incurs an additional phase factor ($e^{\pm i 2 \pi / 3}$).
The total (symmetry-adapted) eigenstate is found by summing the results of all symmetry operations, including these phase factors (Fig.~\ref{Fig::wavefunction}b).
Additionally, a phase factor must also be considered when the atomic orbital itself is rotated, which depends on its azimuthal phase (represented by the color in Fig.~\ref{Fig::wavefunction}a,b).
The conduction band states at the K (K') points are formed mainly by $d_{z^2}$ orbitals, which do not have any azimuthal phase. 
For these eigenstates, the only phase contribution when combining orbitals in different lattice sites arises from the winding of the k-vectors, leading to an out-of-site phase winding.
Thus, the conduction band wavefunctions at the K (K') points transform according to the E'\textsubscript{+} (E'\textsubscript{-}) IR of the point group C\textsubscript{3h}.
In contrast, valence band states are formed predominantly by linear combinations of the $d_{x^2-y^2}$ and $d_{xy}$ orbitals.
These states are combined either as ($d_{x^2-y^2} + i d_{xy}$) or as ($d_{x^2-y^2} - i d_{xy}$), such that they have an orbital angular momentum-like phase winding within the atomic orbital (small arrows in \ref{Fig::wavefunction}b, lower panel).
This is in contrast with the out-of-site phase winding of the conduction band states, which gives implications to the spin-orbit coupling as explained in the following paragraphs.
These linear combinations gain a phase factor of $e^{i 2 \pi / 3}$ or $e^{ -i 2 \pi / 3}$, respectively, when subject to a three-fold rotation.
Combined with the phase acquired due to the winding of the k-vector, this gives rise to a wavefunction composed of a fully in-phase linear combination of orbitals in adjacent lattice sites.
In symmetry terms, these valence band wavefunctions transform as the A' IR of the C\textsubscript{3h} point group.
We note that, to describe the full microscopic character of the VB and CB wavefunctions, we should also consider contributions from the chalcogen orbitals.
We chose not to do this in our approximation since it will not impact the symmetry character of the wavefunctions, although being relevant for quantitatively estimating matrix elements and energy splittings.

\subsection{Symmetries of the spin-orbit coupled eigenstates}

\textbf{7.} Finally, we must also take into account the electron and hole spin when obtaining the symmetry of the wavefuntions.
This can be done by the use of a double group approach \cite{Dresselhaus2008}.
The symmetry of the spin-orbit coupled wavefunction can be obtained by taking the product $\Gamma_{spatial} \times \Gamma_{spin}$, where $\Gamma_{spatial}$ is the IR describing the transformation properties of the spatial wavefunction, and $\Gamma_{spin}$ describes the transformation properties of a spin $1/2$.
A free spin up transforms as the IR $^2\bar{E}_3$ of the double group $\overline{\text{C}}_{\text{3h}}$, whereas a free spin down (its time-reversal conjugate) transforms as IR $^1\bar{E}_3$.
Note here that a rotation by $2 \pi$ adds a phase of $-1$ on the spin $1/2$ state.
Based on this, we can obtain the symmetry properties of the spin-resolved wavefunctions at the edges of the valence and conduction bands at K and K' points, shown in Fig.~\ref{Fig::wavefunction}c.

\textbf{8.} All IRs of the double group $\overline{\text{C}}_{\text{3h}}$ are non-degenerate.
This means that, as has been widely established \cite{Xiao2012, Hsu2015, Song2013, Jiang2017, Xu2014}, spin and valley degrees of freedom are coupled in both valence and conduction bands, giving rise to non-degenerate spin-polarized states.
In this way, spin-up and spin-down states in both valence and conduction bands are split by a spin-orbit energy splitting.
The sign and magnitude of this spin-orbit energy splitting depends on the material properties and cannot be obtained from this purely group-theoretical approach.
Despite the differences between the various TMDs, however, this spin-orbit splitting is in general an order of magnitude larger in the valence band (usually hundreds of meVs) than in the conduction band (usually tens of meVs) \cite{Kormanyos2015}.
We can understand this order-of-magnitude difference based on the considerations above.
For wavefunctions in the valence band, the orbital angular momentum arises from the atomic orbitals themselves, which show an azimuthal phase winding around the transition metal nuclei (as indicated by the color and small arrows in the lower panel of Fig.~\ref{Fig::wavefunction}b). 
This is clear if we rewrite the linear combinations of $d_{x^2-y^2}$ and $d_{xy}$ in terms of spherical harmonics.
This large and well defined orbital angular momentum, localized around the nuclei, gives rise to a large spin-orbit coupling energy.
In contrast, in the conduction band states, there is an intercellular angular momentum arising from phase-winding between different lattice sites \cite{Aivazian2015,Srivastava2015}. 
In addition to that, we note that hybridization with $p$-orbitals also plays a role on the SOC magnitude in the valence band, which is not explicitly considered here.

\textbf{9.} We note that the ordering of states as depicted in Fig.~\ref{Fig::wavefunction}c is valid for tungsten based TMDs; for molybdenum based TMDs, the order in energy of CB1 and CB2 is reversed \cite{Kormanyos2015}.
Nonetheless, the group-theoretical considerations presented here do not depend on the energy ordering of states, and remains valid for both cases.
In what follows, we will focus on the symmetry-restricted scattering processes for charge carriers in the top sub-band of the valence band (VB1, transforming as K\textsubscript{7,8}), and in the two sub-bands of the conduction band (CB1,2 transforming as K\textsubscript{9-12}). 
We disregard the impact of states belonging to the lower sub-band of the valence band (VB2) due to the large SOC energy splitting of hundreds of meV.
Nonetheless, since these states also transform as K\textsubscript{7,8}, this does not incur in any loss of generality since all scattering mechanisms obtained involving VB1 would be the same as the ones involving VB2.

\subsection{Selection rules}

\textbf{10.} Given the symmetries of the various wavefunctions at the band-edges, we can obtain the selection rules governing the spin-flipping scattering processes in these materials at low temperatures. 
According to Fermi's golden rule, a charge carrier in a state $\ket{\psi_i}$ can only scatter into a state $\ket{\psi_f}$ due to a perturbation $H'$ if the matrix element $\braket{\psi_i}{H'|\psi_f}$ is nonzero.
In symmetry terms, this means that the scattering is only possible when the product of IRs $\Gamma_i^* \otimes \Gamma_{H'} \otimes \Gamma_f$ contains the fully symmetric representation, i.e. A'. 
Here, $\Gamma_{i(f)}$ indicates the IR of the initial (final) states, whereas $\Gamma_{H'}$ indicates the IR of the operator responsible for the perturbation. 
Additionally, we note that phonons interact with electrons via the electric fields created by atomic displacements.
In this way, we can also consider selection rules for phonon-driven transitions by looking at the symmetries of these electric fields.
Using this, we can determine which spin-flip scattering processes a perturbation can cause, just by looking at the symmetry of the perturbation.
In the following we focus on the spin flipping mechanisms at the K and K' points of the BZ.
In the supplementary information we provide the product tables and the analysis also for spin-conserving transitions and scattering into other points of the BZ.

\begin{figure}[ht]
  \includegraphics[width=\columnwidth]{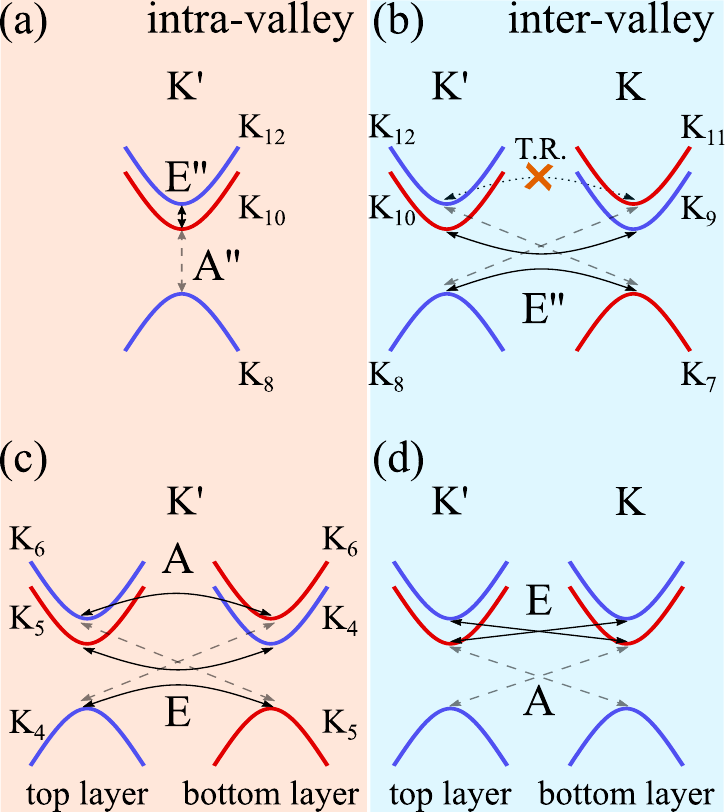}
  \caption{\textbf{Symmetries of spin-flipping scattering mechanisms in ML and BL TMDs.} In monolayer TMDs, only operators transforming as the E" and A" IR of the C\textsubscript{3h} group can lead to either intra-valley (a) or inter-valley (b) spin-flipping scattering processes. In these materials, states transforming as K\textsubscript{11,12} distinguish themselves since energy conserving scattering between these states is forbidden by time-reversal symmetry (see text). In contrast, in bilayer TMDs, additional inter-layer spin-flipping scattering processes arise (c,d), that can couple to external fields transforming as either the A or E IR of the C\textsubscript{3} point group. Here, optical transitions are denoted by a dashed line.
  \label{Fig::SpinFlip}}
\end{figure}

\textbf{11.} These selection rules (for spin-flipping transitions only) are presented comprehensively for a monolayer TMD in Fig.~\ref{Fig::SpinFlip}a,b.
Only operators transforming as A" and E" can generate spin-flipping transitions in ML TMDs. 
In Fig.~\ref{Fig::SpinFlip}, dashed arrows indicate optical transitions. 
These transitions can be actively driven by electro-magnetic fields in the optical spectrum, or arise from radiative electron-hole recombination.
Comparison with Tab.~\ref{tab::ML} shows that spin-flipping direct optical transitions are associated with absorption or emission of electric fields polarized perpendicular to the plane of the TMD layer.
The creation of these so-called 'dark' excitons via this process has been demonstrated by illuminating the TMD monolayer with a parallel beam polarized out-of-plane \cite{Wang2017}.
Additionally, it has been shown that an external in-plane magnetic field can also 'brighten' these transitions, which can be understood as a mixing between the two states belonging to CB1 and CB2 \cite{Zhang2017, Robert2017}.

\begin{table}[h]
	\caption{Symmetries of operators and phonon modes according to the group of the wavevector at the K and K' points for monolayer TMDs. LA (LO), TA (TO) and ZA (ZO) correspond to longitudinal, transverse and out-of-plane acoustic (optical) phonon modes, respectively, \cite{Ribeiro-Soares2014, ZhangX2015}} \label{tab::ML}
	\begin{tabular}{|c|c|c c|c c|}
	\hline
    \multirow{2}{3em}{C\textsubscript{3h} IR} & \multirow{2}{3em}{EM fields} & \multicolumn{2}{|c|}{Acoustic phonons} & \multicolumn{2}{|c|}{Optical phonons}\\
    & & $\Gamma, \mathbf{q}=0$ & $K, \mathbf{q}\neq0$ & $\Gamma, \mathbf{q}=0$ & $K, \mathbf{q}\neq0$\\
    \hline
    & & & & &\\
    A' & $\mathbf{B}_\bot$& & LA/TA & ZO & LO/TO \\
    A'' & $\mathbf{E}_\bot$& ZA &  & ZO & LO/TO\\
	E' & $\mathbf{E}_\parallel$& LA/TA & LA/TA & LO/TO & LO/TO\\
	E'' & $\mathbf{B}_\parallel$& & ZA & LO/TO & LO/TO/ZO\\\hline
	\end{tabular}
\end{table}	

\textbf{12.} Besides this spin-flipping optical transition, only operators transforming as E" of the C\textsubscript{3h} point group can give rise to spin-scattering transitions between electronic states at the various band edges in ML TMDs.
These operators correspond to magnetic fields in the plane of the ML, out-of-plane phonons at the K-point of the acoustic phonon band and optical phonons at both $\Gamma$- and K-points, for example (see Tab.~\ref{sec::ML}). 
These phonon modes have energies on the order of hundreds of meVs \cite{Ribeiro-Soares2014, ZhangX2015}. 
Thus, they will be very weakly populated at cryogenic temperatures, leading to a suppression of phonon-related upwards scattering processes. 
We note that, this is valid also for Mo-based TMDs, despite their small SOC splitting between CB1 and CB2. 
In this case, even though the bands are close in energy, hot phonons are required to drive the transition.
In contrast, downwards scattering processes (K$_{12}$ $\rightarrow$ K$_{10}$) can happen via the emission of a phonon even at low temperatures.
This process should be distinct for molybdenum and tungsten based TMDs, since the order of the two bands are interchanged, while the optical transition used to generate a spin-valley population (K$_{8}$ $\rightarrow$ K$_{12}$) is the same.
Our analysis then indicates that, at low temperatures, the spin lifetime for electrons in molybdenum-based TMDs should be in principle longer than when compared to tungsten-based ones.
This can be used to understand the long spin-lifetimes recently reported for MoSe$_2$, which persist up to room temperature \cite{Ersfeld2019}.
We do note that in-plane magnetic fields will be more effective in driving intra-valley spin-flips of spins in the CB of Mo-based TMDs, when compared to W-based TMDs.
This is because the energy splitting between CB1 and CB2 is much smaller in the former, and thus easier to overcome by Zeeman-like energy terms.


\textbf{13.} When considering inter-valley scattering processes, we must note that additional selection rules arise due to time-reversal symmetry and Kramer's theorem (see Supplementary Material).
Kramer's theorem states that, if time-reversal symmetry is preserved, wavefunctions connected by conjugation have the same energy, that is, a spin-up in a K valley and the corresponding spin-down in the K' valley have the same energy.
This implies that energy-conserving spin-flipping scattering processes between the K and K' valleys (inter-valley) can only arise due to perturbations that break time-reversal symmetry (see supplementary information).
Transitions between time-conjugate pairs transforming as $\text{K}_7 \leftrightarrow \text{K}_8$ and $\text{K}_9 \leftrightarrow \text{K}_{10}$ can arise due to perturbations transforming as E".
Since in-plane magnetic fields and K-phonons transforming as E" break time-reversal symmetry, these scattering processes are thus fully allowed, also by time-reversal (TR) symmetry \cite{ZhangL2015}.
This is in line with the experimental observations that identify out-of-plane K-phonons as the main sources of hole spin-valley depolarization in both W and Mo-based TMDs \cite{Hsu2015, Mak2012, Zeng2012}.
In contrast, transitions between states transforming as $\text{K}_{11} \leftrightarrow \text{K}_{12}$ are allowed -- considering only spatial symmetry -- when these states interact with external fields transforming as A".
Out-of-plane electric fields and $\Gamma$-point phonons (see Tab.~\ref{tab::ML}) preserve time-reversal symmetry, such that they cannot drive these transitions.
In contrast, chiral K-phonons in the optical phonon bands \cite{ZhangL2015} do break time-reversal symmetry and could drive these transitions. Nonetheless, these are high-energy phonon modes which are not populated at low temperatures.

\textbf{14.} The results of the last paragraph point to a fundamental asymmetry in the behavior of electron and hole spin scattering processes in these materials, and are in line with existing literature.
For example, the selection rules obtained above provide an intuitive interpretation of theoretical and experimental results showing that, in Mo-based TMDs, phonon-related spin decay affects holes in VB1 much more efficiently than electrons in CB1  at low temperatures \cite{Ersfeld2019} (note that in Mo-based TMDs, states in CB1 transform as K\textsubscript{11,12}).
Additionally, despite the seemingly simplistic single-particle picture presented here, these results also provide an explanation for the recently observed asymmetry between bright and dark excitons concerning direct to indirect exciton scattering in W-based TMDs \cite{Wu2020}, where indirect excitons are composed of an electron and a hole in opposite valleys.


\textbf{15.} Finally, the selection rules derived in this section imply that, in monolayer TMDs, spin scattering is relatively robust with respect to the presence of noisy electric fields, such as randomly-distributed Coulomb scatterers and local strain.
Only out-of-plane electric fields can cause spin-flipping scattering transitions; however, these transitions are either in the optical range -- such that they must be actively driven or arise from electron-hole radiative recombination -- or forbidden by the requirements of Kramer's theorem.
Thus, in the low energy scattering regime, local (due to the electrostatic environment, strain of the material, substrate effects, etc.) or global electric fields will have limited influence on the prevalence of various spin-scattering processes.
This also means that spin scattering rates in ML TMDs should not be greatly influenced by the symmetry breaking created by a particular substrate.
In contrast, in-plane magnetic fields, either extrinsic or intrinsic to the sample due spin-active defects or nuclear spins, can cause both intra- and inter-valley spin flips. 
On the one hand, these results indicate that the spin-lifetimes in ML devices can be enhanced by ensuring a low concentration of deep-level spin-active lattice defects.
On the other hand, they also indicate that control over the spin-polarization in these materials relies on externally applying magnetic fields, which is a slow and practically challenging process.


\section{Bilayer TMDs}

\subsection{Symmetries and eigenstates}

\textbf{16.} Figures \ref{Fig::geom}c,d show the symmetries of bilayer 2H-TMDs in both real and reciprocal space (symmetry operations valid at the $\Gamma$ point but absent at K and K' points are shown in gray).
When compared to monolayer crystals, bilayer stacks of 2H-TMDs have some notable symmetry changes (see Fig.~\ref{Fig::geom}) \cite{Ribeiro-Soares2014}.
In particular, bilayers lack a horizontal mirror plane, but do have an inversion point which brings the top layer into the bottom one.
The presence of inversion symmetry means that, if we consider the entire stack, spin-valley coupling is not allowed to exist -- electronic eigenstates at the K and K' points are spin degenerate.
However, a local spin polarization of the bands may arise when inversion symmetry is present at a global scale, but is locally broken \cite{Zhang2014}.
Since interlayer coupling is small compared to the other intrinsic energy scales in 2H-TMDs \cite{Gong2013, Jiang2017}, this local spin polarization arises within each ML making up the multilayer stacks, giving rise to a spin-valley-layer coupling \cite{Gong2013, Zhang2014} which has been experimentally observed \cite{Guimaraes2018, Riley2014, Ye2019, Li2020}.
These results are evidence that the layers are partially decoupled, such that the system can be approximately described by a stack of two distinguishable monolayers whose crystallographic symmetry corresponds to point-group C\textsubscript{3v}.
The horizontal mirror plane and two-fold axes in Fig.~\ref{Fig::geom}a,b are not valid symmetry operations anymore, since the top and bottom environments of each layer differ.
At the K and K' points, this symmetry is reduced to C\textsubscript{3}, such that the electronic eigenstates at the edges of the bands in each layer transform as IRs of the double group $\overline{\text{C}}_{\text{3}}$.
Since the bottom layer is inverted with respect to the top layer, the direction of phase winding of the eigenstates at the K and K' points of the BZ happens in opposite directions for each of the layers.
This means that, at a given energy and at a given point of the BZ, eigenstates in different layers will have opposite orbital and spin angular momentum.
This results in a alternating spin-valley ordering according to the layer number, i.e. the top of the valence band of the valley K of one layer has the same spin (and symmetry) as the top of the valence band of the opposite valley (K') of the adjacent layer.
The resulting band structure, with the respective symmetries of each of the eigenstates, can be found in the supplementary information.

\subsection{Selection Rules}

\textbf{17.} Treating the layers as distinguishable does not mean that they are fully independent.
This means that the additional layer degree-of-freedom of TMD bilayers allows for additional inter-layer scattering processes.
The intra-layer scattering processes are the same as the ones treated in detail in Sec. \ref{sec::ML} and will not be repeated here.
The selection rules for the inter-layer processes can be obtained in the same manner as before, now considering eigenstates and operators transforming as IRs of the double group $\overline{\text{C}}_{\text{3}}$.
These additional spin-flipping scattering pathways are shown in Fig.~\ref{Fig::SpinFlip}c,d. 

\textbf{18.} Notably, the situation is drastically different for spins that are protected from energy-conserving spin-flipping scattering processes in a ML due to TR symmetry.
These states transform as K\textsubscript{11,12} in a ML, and as K\textsubscript{6} in the bilayer.
These charge carriers can now flip their spin by going from one layer into the other after interacting with an operator transforming as the A IR of the point group C\textsubscript{3}.
This is because, for states in the K valley for example, a spin-up in the top layer and a spin-down in the bottom layer are not TR conjugates of each other, such that this scattering is not protected by TR symmetry.
These scattering processes can arise from electro-magnetic fields perpendicular to the layer plane, or due to out-of-plane acoustic phonons, which enhance the interlayer coupling (see Tab.~\ref{tab::BL}).
The availability of acoustic phonons at the $\Gamma$ point at low temperatures and the presence of environmental charge noise implies that spins in CB2 (CB1) states in W based (Mo based) TMDs will suffer from significantly faster relaxation than their counterparts in ML TMDs.

\begin{table}[ht]
	\caption{Symmetries of operators and phonon modes according to the group of the wavevector at the K and K' points for bilayer TMDs.} \label{tab::BL}
	\begin{tabular}{|c|c|c c|c c|}
	\hline
    \multirow{2}{3em}{C\textsubscript{3} IR} & \multirow{2}{3em}{EM fields} & \multicolumn{2}{|c|}{Acoustic phonons} & \multicolumn{2}{|c|}{Optical phonons}\\
    & & $\Gamma, \mathbf{q}=0$ & $K, \mathbf{q}\neq0$ & $\Gamma, \mathbf{q}=0$ & $K, \mathbf{q}\neq0$\\
    \hline
    & & & & & \\
    A & $\mathbf{E}_\bot$, $\mathbf{B}_\bot$ & ZA & LA/TA & ZO & LO/TO\\
	E & $\mathbf{E}_\parallel$, $\mathbf{B}_\parallel$& LA/TA & LA/TA/ZA & LO/TO & LO/TO/ZO\\ \hline
	\end{tabular}
\end{table}	

\textbf{19.} Additional inter-layer energy conserving spin-flipping scattering processes also arise for states transforming as K\textsubscript{4,5}.
These processes must be driven by operators transforming as the E IR of the point group C\textsubscript{3}, corresponding to electromagnetic fields in the layer plane, and longitudinal and transverse acoustic phonon modes in the $\Gamma$ point (Tab.~\ref{tab::BL}).
Again, these processes are expected to be prevalent even at low temperatures, leading to fast spin relaxation.

\textbf{20.} Finally, additional inter-layer spin-flipping processes that modify the linear momentum of charge carriers, \textit{i.e.} inter-layer inter-valley processes (Fig.~\ref{Fig::SpinFlip}d), also arise.
Due to requirements of momentum conservation, these processes must be accompanied by the emission or absorption of a K-phonon.
Since they involve a change in the energy of the charge carriers of at least a few tens of meVs, upward scattering processes (CB1 $\rightarrow$ CB2) are likely suppressed at low temperatures.
Relaxation of hot carriers accompanied by a spin flip and change in linear momentum (CB2 $\rightarrow$ CB1) however may arise via the emission of K-phonons transforming as E.
Additionally, in-plane momentum transfer from the the CB at the K point of one layer into the VB at the K' point of the other layer (transitions denoted by dashed lines in Fig.~\ref{Fig::SpinFlip}d), are only allowed by second-order processes involving a photon and a phonon, transforming as the IR A of group C$_{3}$.
Therefore, these transitions should be suppressed at low temperatures.

\textbf{21.} The considerations in the past paragraphs imply that inter-layer scattering processes in bilayers (or few-layer stacks) lead to additional spin relaxation channels, hindering their application in the field of spintronics.
However, they also imply that, in these materials, we have additional control over spin-flipping processes.
Kerr rotation experiments in W based bulk TMDs show that the spin polarization in these materials decays within tens of ps \cite{Guimaraes2018}.
Although much shorter than their counterparts in ML samples \cite{Jones2014, Song2016, Volmer2017}, these spin lifetimes still enable optical detection with high resolution.
The group-theoretical results of this section indicate that externally applied electric fields, for example, could be used to manipulate the spin scattering rates in these materials, allowing us to turn these optically induced spin signals on/off electrically \cite{Gong2013, Jones2014}.
Furthermore, in-plane and out-of-plane electric fields studies could unravel the charge character of the spin polarization.
This results from the fact that in-plane electric fields will impact the spin-scattering processes of optically created holes, whereas out-of-plane electric fields will impact the scattering processes of optically created electrons.

\textbf{22.} Several approaches could be used to enhance the spin lifetimes in bilayer TMDs.
On the one hand, we expect that encapsulating bilayer and bulk devices with van der Waals insulators like hBN will be important in reducing the electrostatic influence of substrate or adsorbed charges.
The electrostatic environment of the BL could be further controlled by gating, possibly leading to longer spin lifetimes.
On the other hand, heterostructures of TMDs will have different interlayer coupling and interfacial electric dipoles, depending on the particular combination of materials.
Engineering these parameters, for example, could significantly suppress interlayer scattering mechanisms, leading to spin lifetimes more similar to those found in ML TMDs \cite{Rivera2018}.
Additionally, the phonon spectra can also be modified by strain or the coupling to other van der Waals materials in heterostructures.
In these devices, one could think of deliberately enhancing interlayer scattering processes via applied electric fields, combining the long spin lifetimes observed in ML TMDs with the enhanced electrical control of spin polarization provided by multilayer stacks.

\textbf{23.} Finally, we note that optical fields, such as circularly polarized light, will couple to a certain spin species according to the selection rules established above.
When one applies an additional static electric or magnetic field that induces state mixing between different layers, this picture still does not change, \textit{i.e.} circularly polarized light will still couple to the same spin species.
However, when the optical field is turned off, the spin polarization will evolve according to the coupled Hamiltonian given by the perturbing static electric or magnetic field.
This leads to an oscillation between the two states in time, reminiscent of what is observed in experiments studying the coherent evolution of optically created spins in III-V and II-VI semicondictors, for example \cite{Kikkawa1997,Kimel2001}. These (Rabi) oscillations will decay according to the characteristic spin relaxation and dephasing times \cite{Gong2013}.

\section{Conclusion}

\textbf{24.} Group theory is a powerful tool in the analysis of both equilibrium and out-of-equilibrium physical processes in a variety of materials, including TMDs.
It allows one to gain insight into complex physical phenomena from a mathematically simple and comprehensive tool, without needing the specifics of the material of interest.
Additionally, it allows one to broadly generalize insights obtained for one material or set of electronic eigenstates without additional computational cost.
This approach is simplistic, and relies on a series of approximations. Nonetheless, we have shown here that it helps to unveil the fundamental processes at play in various experiments on TMDs.
Even the behavior of excitons -- for which, the presence of exchange interaction is not encompassed by the single-particle approach undertaken here -- can often be explained qualitatively by this group-theoretical model, by treating the electron and hole separately \cite{Robert2017}.

\textbf{25.} Based on this group-theoretical approach, we could identify fundamental symmetry properties of spins in TMDs and the subsequent selection rules for spin-scattering processes.
In ML TMDs, charge carriers in each of the sub-bands of the CB behave in a fundamentally different manner: for one of the CB sub-bands, energy conserving spin-flipping processes are forbidden by TR symmetry, suppressing most of the phonon-related spin-flips at low temperatures.
This is not true for the other sub-band of the CB, and for the VB, such that charge carriers in these states can have their spin flipped via scattering from a K-phonon.
Thus, in Mo-based TMDs -- where the states that are symmetry protected with regards to spin-flips sit at the bottom of the CB and the SOC splitting in the CB is smaller -- magnetic impurities should be the main source of spin-flipping scattering at low temperatures.
In these materials, the quality of the sample can drastically enhance the spin lifetime of electrons in the edge of the CB, possibly explaining the broad variation of spin lifetimes reported in literature.
Additionally, we find that spin-scattering in ML TMDs is very robust with respect to electric fields, with only fields in the optical range actually giving rise to spin-flips, and for a restricted set of states.
In contrast, in BL TMDs, all electronic states can undergo spin-flips after interacting with electric fields that cause interlayer momentum-conserving transitions.
Thus, noisy electric fields, an inhomogeneous electrostatic environment, and acoustic phonons are expected to greatly suppress a spin polarization induced in these materials, and decrease their lifetime.
Nonetheless, this feature can also be harnessed to gain control over optically created and detected spin polarization in these materials via electrostatic gating, for example.
There, one can use an out-of-plane electric field to efficiently control the spin relaxation in these materials, making it a viable option for spin-based information processing.

\section{Acknowledgements}

This work was supported by the Zernike Institute for Advanced Materials and the Dutch Research Council (NWO, STU.019.014).

\bibliography{myBib}

\end{document}


%

\setcounter{page}{1}
\setcounter{tocdepth}{1}
\renewcommand{\thefigure}{S\arabic{figure}}
\setcounter{figure}{0}
\renewcommand{\thetable}{S\Roman{table}}
\setcounter{table}{0}


\newcommand{\be}[1]{\begin{eqnarray}  {\label{#1}}}
\newcommand{\ee}{\end{eqnarray}}

\begin{center}
\textbf{{\LARGE Supplementary Information}}

for

\textbf{{\Large Symmetry Evolution and Control of Spin-Scattering Processes in \\Two-Dimensional Transition Metal Dichalcogenides}}

by

Carmem~M.~Gilardoni, Freddie~Hendriks, Caspar~H.~van~der~Wal and Marcos~H.~D.~Guimar\~aes

\emph{Version of \today}

\end{center}

\vspace{4cm}

\tableofcontents


\newpage

\section{Electronic eigenstates and selection rules for monolayer TMDs}

Here, we present the derivation of the symmetry of the electronic eigenstates at the high symmetry points of the Brillouin Zone (BZ), and the subsequent symmetry restricted selection rules for spin scattering processes for the monolayer. 
The symmetries of the group of the wave vector at various points of the BZ were previously derived and can be found also in literature \cite{Ribeiro-Soares2014}. 
These symmetries are presented in more detail and can be visualized in Fig.~I of the main text. 
At the $\Gamma$ point of the BZ, wavefunctions have the symmetries of the point group D\textsubscript{3h}. At the K and K' points, this symmetry is reduced to C\textsubscript{3h}.
At the Q and Q' points, it is further reduced to C\textsubscript{s}\textsuperscript{xy} (also denoted by C\textsubscript{1h}).

In order to account for the transformation properties of spin-half states, we make use of the double groups with the symmetries mentioned above \cite{Dresselhaus2008}.
The character tables referring to these double groups are shown explicitly in section \ref{subsec::ML_Char}.
In Sec. \ref{subsec::ML_SymWF}, we obtain the symmetries of the electronic eigenstates at the high symmetry points of the BZ based on the symmetry of the crystal ($\Gamma_\text{equiv}$) and of the atomic orbitals most prevalent in each of the eigenstates ($\Gamma_\text{orb}$). 
Finally, in Sec. \ref{subsec::ML_Selection}, we use the symmetries of these eigenstates to obtain the selection rules regarding spin-scattering processes. 

\subsection{Character tables}
\label{subsec::ML_Char}

Here, we provide the character tables for the double groups of interest. We denote the irreducible representations (IR) by the point of the BZ associated with each particular group ($\Gamma$, K and Q). In parenthesis, we also provide the Mulliken symbol associated with each IR. The IRs indicating the transformation properties of spin 1/2 states (or spin $N/2$, with odd $N$) are indicated by an overbar. In the following character tables, when present the symbol $\omega$ corresponds to $\exp{2\pi i/3}$, whereas $\omega^*$ corresponds to $\exp{-2\pi i/3}$.

\begin{table}[h]
	\begin{tabular}{|l|c c c c c c c c c|}
		\hline
		D\textsubscript{3h} & $E$ & $\bar{E}$ & $C_3^+ C_3^-$ & $\bar{C}_3^+ \bar{C}_3^-$ & $\sigma_h \bar{\sigma}_h$ & $S_3^+ S_3^-$ & $\bar{S}_3^+ \bar{S}_3^-$ & $ C_{2i}' \bar{C}_{2i}'$ & $\sigma_{vi} \bar{\sigma}_{vi}$ \\ \hline
		
		$\Gamma_1$ $(\text{A}_1')$ &1 &1 &1 &1 &1 &1 &1 &1 &1 \\
		$\Gamma_2$ $(\text{A}_2')$ &1 &1 &1 &1 &1 &1 &1 &-1 &-1 \\
		$\Gamma_3$ $(\text{A}_1'')$ &1 &1 &1 &1 &-1 &-1 &-1 &1 &-1 \\
		$\Gamma_4$ $(\text{A}_2'')$ &1 &1 &1 &1 &-1 &-1 &-1 &-1 &1 \\
		$\Gamma_5$ $(\text{E}'')$ &2 &2 &-1 &-1 &-2 &1 &1 &0 &0 \\
		$\Gamma_6$ $(\text{E}')$ &2 &2 &-1 &-1 &2 &-1 &-1 &0 &0 \\ \hline
		$\Gamma_7$ $(\bar{\text{E}}_1)$ &2 &-2 &1 &-1 &0 &$\sqrt{3}$ &$-\sqrt{3}$ &0 &0 \\
		$\Gamma_8$ $(\bar{\text{E}}_2)$ &2 &-2 &1 &-1 &0 &$-\sqrt{3}$ &$\sqrt{3}$ &0 &0 \\
		$\Gamma_9$ $(\bar{\text{E}}_3)$ &2 &-2 &-2 &2 &0 &0 &0 &0 &0 \\ \hline
	\end{tabular}
\end{table}

\begin{table}[h]
	\begin{tabular}{|l|c c c c c c c c c c c c|}
		\hline
		C\textsubscript{3h} & $E$ & $C_3^+$ &  $C_3^-$ & $\sigma_h$ & $S_3^+$ & $S_3^-$ & $\bar{E}$ & $\bar{C}_3^+$ & $\bar{C}_3^-$ & $\bar{\sigma}_h$ & $\bar{S}_3^+$ & $\bar{S}_3^-$ \\ \hline
		
		$\text{K}_1$ $(\text{A}')$ &1 &1 &1 &1 &1 &1 &1 &1 &1 &1 &1 &1 \\
		$\text{K}_2$ $(^2\text{E}')$ &1 &$\omega$ &$\omega^*$ &1 &$\omega$ &$\omega^*$ &1 &$\omega$ &$\omega^*$ &1 &$\omega$ &$\omega^*$ \\
		$\text{K}_3$ $(^1\text{E}')$ &1 &$\omega^*$ &$\omega$ &1 &$\omega^*$ &$\omega$ &1 &$\omega^*$ &$\omega$ &1 &$\omega^*$ &$\omega$ \\
		$\text{K}_4$ $(\text{A}'')$ &1 &1 &1 &-1 &-1 &-1 &1 &1 &1 &-1 &-1 &-1 \\
		$\text{K}_5$ $(^2\text{E}'')$ &1 &$\omega$ &$\omega^*$ &-1 &$-\omega$ &$-\omega^*$ &1 &$\omega$ &$\omega^*$ &-1 &$-\omega$ &$-\omega^*$ \\
		$\text{K}_6$ $(^1\text{E}'')$ &1 &$\omega^*$ &$\omega$ &-1 &$-\omega^*$ &$-\omega$ &1 &$\omega^*$ &$\omega$ &-1 &$-\omega^*$ &$-\omega$ \\ \hline
		$\text{K}_7$ $(^2\bar{\text{E}}_3)$ &1 &$-\omega$ &$-\omega^*$ &$i$ &$-i\omega$ &$i\omega^*$ &-1 &$\omega$ &$\omega^*$ &$-i$ &$i\omega$ &$-i\omega^*$ \\
		$\text{K}_8$ $(^1\bar{\text{E}}_3)$ &1 &$-\omega^*$ &$-\omega$ &$-i$ &$i\omega^*$ &$-i\omega$ &-1 &$\omega^*$ &$\omega$ &$i$ &$-i\omega^*$ &$i\omega$ \\
		$\text{K}_9$ $(^2\bar{\text{E}}_2)$ &1 &$-\omega$ &$-\omega^*$ &$-i$ &$i\omega$ &$-i\omega^*$ &-1 &$\omega$ &$\omega^*$ &$i$ &$-i\omega$ &$i\omega^*$ \\
		$\text{K}_{10}$ $(^1\bar{\text{E}}_2)$ &1 &$-\omega^*$ &$-\omega$ &$i$ &$-i\omega^*$ &$i\omega$ &-1 &$\omega^*$ &$\omega$ &$-i$ &$i\omega^*$ &$-i\omega$ \\		
		$\text{K}_{11}$ $(^2\bar{\text{E}}_1)$ &1 &-1 &-1 &$i$ &$-i$ &$i$ &-1 &1 &1 &$-i$ &$i$ &$-i$ \\
		$\text{K}_{12}$ $(^1\bar{\text{E}}_1)$ &1 &-1 &-1 &$-i$ &$i$ &$-i$ &-1 &1 &1 &$i$ &$-i$ &$i$ \\ \hline
	\end{tabular}
\end{table} 

\begin{table}[h]
	\begin{tabular}{|l|c c c c|}
		\hline
		C\textsubscript{1h} & $E$ & $\sigma_h$ & $\bar{E}$ & $\bar{\sigma}_h$ \\ \hline
		$\text{Q}_1$ $(\text{A}')$ &1 &1 &1 &1\\
		$\text{Q}_2$ $(\text{A}'')$ &1 &-1 &1 &-1\\ \hline
		$\text{Q}_3$ $(^1\bar{\text{E}})$ &1 &$i$ &-1 &$-i$\\
		$\text{Q}_4$ $(^2\bar{\text{E}})$ &1 &$-i$ &-1 &$i$\\ \hline
	\end{tabular}
\end{table}


\FloatBarrier

\subsection{Band structure and symmetry}
\label{subsec::ML_SymWF}

As stated in the main text, the atomic orbital character of the electronic eigenstates in the high symmetry points of the BZ has been obtained in literature based on both tight-binding and density functional theory models \cite{Song2013, Liu2013, Kormanyos2015}. Table \ref{TabSI::ML_GWV} presents, for the three high symmetry points of the BZ, the point-group associated with the group of the wave vector (GWV) in column PG (for point-group), the atomic orbital character and its symmetry ($\Gamma_\text{orb}$) and the symmetry of the equivalent representation ($\Gamma_\text{equiv}$). Combined, this information allows us to obtain the symmetry of the delocalized Bloch wavefunction at each high-symmetry point of the BZ, given by $\Gamma_\text{spatial}$. For the states in the K and K' points, this derivation is performed visually and explained thoroughly in the main text. For the valence band states in the K and K' points, the 2-fold degenerate atomic orbitals, combined with the 2-fold degenerate equivalent representation gives rise to a total of 4 spatial wavefunctions. Nonetheless, only those transforming as A' are of importance here, since the other ones (identified in parenthesis in the last column of Tab.~\ref{TabSI::ML_GWV}) are much lower in energy, deep into the VB. 

\begin{table}[]
    \caption{Spatial character of the wavefunctions at edges of CB and VB at various points of the BZ of monolayer TMDs and their symmetries.}
    \label{TabSI::ML_GWV}
	\begin{tabular}{|c|c|c|c|c|c|}
		\hline
			BZ point & PG & Atomic Orbital & $\Gamma_{\text{orb}}$ & $\Gamma_\text{equiv}$ & $\Gamma_\text{spatial} = \Gamma_\text{equiv} \otimes \Gamma_{\text{orb}}$ \\ \hline
			
			$\Gamma$, VB & D\textsubscript{3h} &  d\textsubscript{z\textsuperscript{2}} & $\text{A}'_{1}$ & $2 \text{A}'_{1}$ & $2 \text{A}'_{1}$ \\ \hline
						
			K (\text{K}'), VB & C\textsubscript{3h} & d\textsubscript{xy, x\textsuperscript{2}-y\textsuperscript{2}} &  $\text{E}'_{+} + \text{E}'_{-}$ & $\text{E}'_{+} + \text{E}'_{-}$ & $2 \text{A}' + (\text{E}'_{+} + \text{E}'_{-})$ \\
			
			K (\text{K}'), CB & C\textsubscript{3h} & d\textsubscript{z\textsuperscript{2}} & $\text{A}'$ & $\text{E}'_{+} + \text{E}'_{-}$ & $\text{E}'_{+} + \text{E}'_{-}$ \\  \hline
			
			Q (Q'), CB & C\textsubscript{s}\textsuperscript{xy} & d\textsubscript{xy, x\textsuperscript{2}-y\textsuperscript{2}} & $2 \text{A}'$ & $\text{A}'$ & $2\text{A}'$ \\ \hline
	\end{tabular}
\end{table}

The double groups $\bar{\text{C}}_{\text{3h}}$ and $\bar{\text{C}}_{\text{1h}}$ do not have any IR of dimension 2 or greater. This means that all eigenstates in the K (K') or Q (Q') of the BZ are non-degenerate, even when spin is included. 
Degeneracy here is restricted to cases where two states have the same energy at the same point of the BZ (thus, states with the same energy but at $K$ and $\text{K}'$ points are not considered degenerate).
In contrast, at the $\Gamma$ point, spin up and down states must be degenerate in the presence of time-reversal symmetry, since all spin 1/2 IRs of the group D\textsubscript{3h} are two-fold degenerate. 

\begin{figure}
	\centering
	\includegraphics[width=\linewidth]{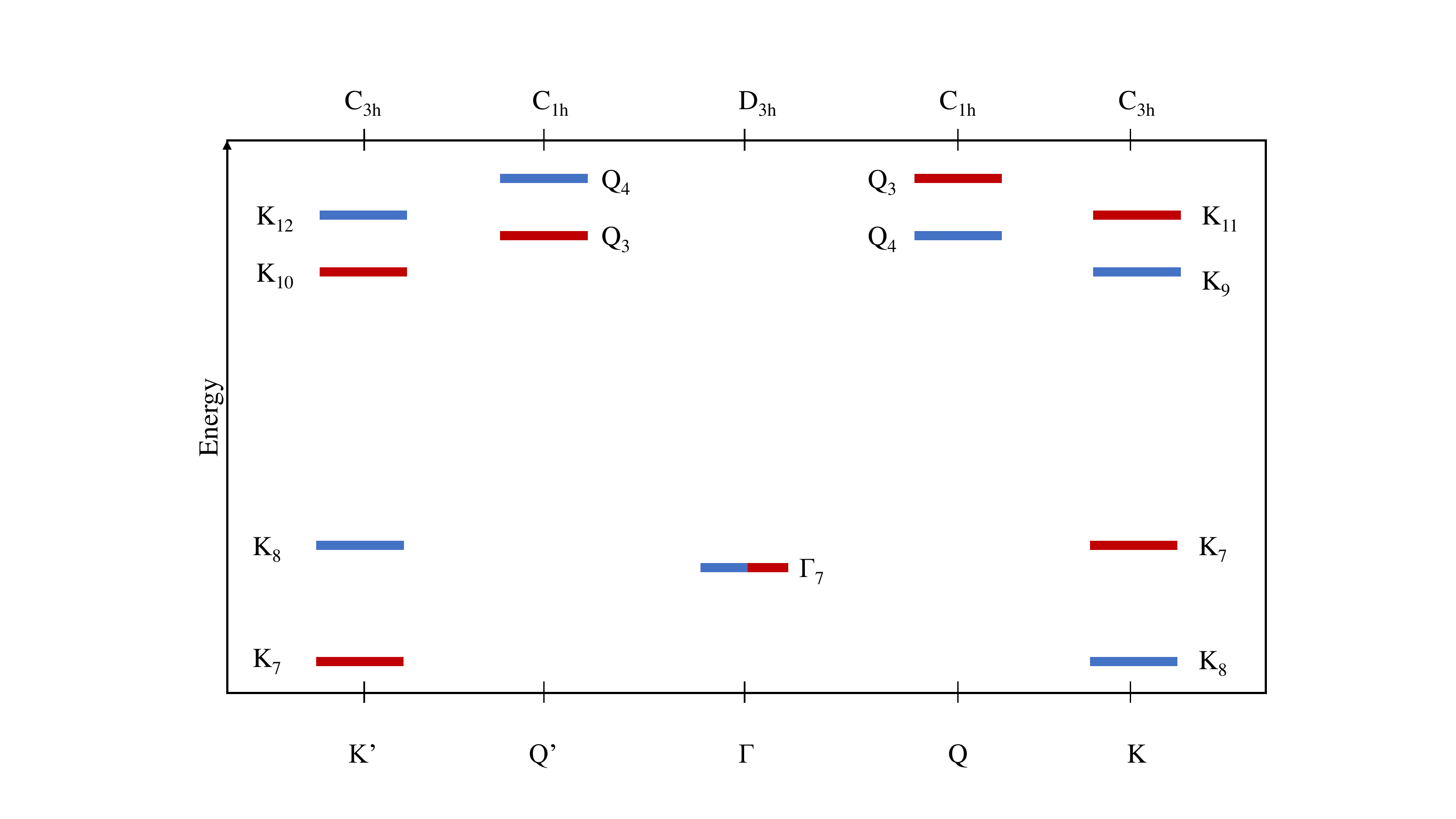}\\
	\caption{\textbf{Energy diagram ML} Blue lines represent spin down states, whereas red lines represent spin up states. Lines of both colors represent degeneracies protected by symmetry. The IRs associated with each state are those of the group of the wave vector at each point of the BZ, indicated at the top.}
	\label{figSI::ML}
\end{figure}

In tables \ref{TabSI::ML_VB} and \ref{TabSI::ML_CB} we present, respectively for valence and conduction band states, the symmetries of the electronic eigenstates including spin. 
This is given in column $\Gamma_\text{spatial} \otimes \Gamma_\text{spin}$. 
This information is also compiled visually in Fig.~\ref{figSI::ML}.
In order to obtain the selection rules between states at different points in the BZ (which have different symmetries), we must take into account the compatibility relations between different point groups \cite{Dresselhaus2008}. 
These compatibility relations state how an IR of a certain point group splits into a sum of IRs of a different group upon lowering the symmetry. 
In tables \ref{TabSI::ML_VB} and \ref{TabSI::ML_CB}, we also provide the compatibility relations for the IRs corresponding to the electronic eigenstates at each high symmetry point of the BZ, for all three point groups of interest. 

\begin{table}[h]
	\caption{For valence band wavefunctions, we can obtain the symmetry of the electronic eigenstates including spin by considering the IRs of the double groups. We present also the compatibility relations for IRs of the double groups D\textsubscript{3h}, C\textsubscript{3h} and C\textsubscript{1h}.}
	\label{TabSI::ML_VB}
	\begin{tabular}{|c|c|c|c c c|}
		\hline
		BZ point & VB wf & $\Gamma_{\text{spatial}} \otimes \Gamma_{\text{spin}}$ & D\textsubscript{3h} & C\textsubscript{3h} & C\textsubscript{1h} \\ \hline
		$\Gamma$ & $\text{A}_1' \otimes \text{E}_{1/2}$ & $\Gamma_7$ & & $\text{K}_7 + \text{K}_8$ & $\text{Q}_3 + \text{Q}_4$ \\
		K (\text{K}') & $\text{A}' \otimes \text{E}_{1/2 \uparrow}$ & $\text{K}_{7}$ & $\Gamma_7$ & & $\text{Q}_3$ \\ 
		& $\text{A}' \otimes \text{E}_{1/2 \downarrow}$ & $\text{K}_{8}$ &  & & $\text{Q}_4$ \\ \hline
	\end{tabular}
	
\end{table}

\begin{table}[h]
\caption{For conduction band wavefunctions, we can obtain the symmetry of the electronic eigenstates including spin by considering the IRs of the double groups. We present also the compatibility relations for IRs of the double groups D\textsubscript{3h}, C\textsubscript{3h} and C\textsubscript{1h}.}
\label{TabSI::ML_CB}
	\begin{tabular}{|c|c|c|c c c|}
		\hline
		BZ point & CB wf & $\Gamma_{\text{orb}} \otimes \Gamma_{\text{spin}}$ & D\textsubscript{3h} & C\textsubscript{3h} & C\textsubscript{1h}\\ \hline
		T (T') & $2 \text{A}' \otimes \text{E}_{1/2 \uparrow}$ & $2 \text{Q}_3$ & & &\\
		 & $2 \text{A}' \otimes \text{E}_{1/2 \downarrow}$ & $2 \text{Q}_4$ & & &\\
		$\text{K} (\text{K}')$ & $\text{E}'_{+} \otimes \text{E}_{1/2 \uparrow}$ & $\text{K}_{10}$ & $\Gamma_8$ & & $\text{Q}_3$\\
		 & $\text{E}'_{-} \otimes \text{E}_{1/2 \downarrow}$ & $\text{K}_{9}$ & & & $\text{Q}_4$\\
		 & $\text{E}'_{+} \otimes \text{E}_{1/2 \downarrow}$ & $\text{K}_{12}$ & $\Gamma_9$ & & $\text{Q}_4$\\
     	 & $\text{E}'_{-} \otimes \text{E}_{1/2 \uparrow}$ & $\text{K}_{11}$ & & & $\text{Q}_3$\\ \hline
	\end{tabular}
	
\end{table}

\FloatBarrier

\subsection{Possible scattering processes}
\label{subsec::ML_Selection}

According to Fermi's golden rule, the probability of going from an initial state $\ket{\Psi_i}$ to a final state $\Psi_f$ when interacting with a perturbation $H'$ is proportional to the matrix element $\braket{\Psi_f}{H'|\Psi_i}$. 
This matrix element must be a scalar, that is, it must transform as the fully symmetric representation A (or A\textsubscript{$1g$}). 
Thus, the matrix element can only be nonzero if the product of the representations $\Gamma_f^* \otimes \Gamma_{H'} \otimes \Gamma_i$, respectively indicating the symmetry properties of $\ket{\Psi_f}$, $H'$ and $\ket{\Psi_i}$, contains the fully symmetric representation. 
When states $\ket{\Psi_i}$ and $\ket{\Psi_f}$ occur at different points in the BZ, we must consider their symmetry properties according to the point group of lowest symmetry by use of the previously mentioned compatibility relations.
In this way, we can calculate all matrix elements which may be nonzero. 
We do so, and compile the results in table \ref{TabSI::ML_Selection}. 
Here, we classify the transitions with respect to whether they preserve spin or valley information.
Furthermore, we identify the external electromagnetic field component which transforms according to each of the IRs and is, thus, capable of driving a certain transition. 
Optical transitions are highlighted in red. 
We highlight these transitions since they involve a large change in the energy of the electronic eigenstate and can be addressed by light. 
Thus, at low temperatures, these transitions will only happen upwards if actively driven (the downwards transition can happen via spontaneous emission of a photon). 
Furthermore, we did not present the selection rules for optical transitions that do not conserve linear momentum. These transitions happen in two steps, via the emission of a photon and a phonon, such that these processes are already considered here. 

Finally, we highlight transitions between states that are each other's time-reversal conjugate in blue. 
Transitions between these states are further protected by time reversal symmetry, according to Kramer's Theorem. 
If two states $\ket{\Psi_i}$ and $\ket{\Psi_f}$ are each other's time reversal conjugate (and they are both spin-1/2 states), the matrix element $\braket{\Psi_f}{H'|\Psi_i}$ is nonzero only if the perturbation $H'$ breaks time reversal symmetry.
Examples of such perturbations are external magnetic fields, magnetic scatterers and circularly polarized light. 

\begin{table}[h]
\caption{Scattering mechanisms for electrons in K and $\text{K}'$ points in monolayer TMDs. The abbreviations Magn. and El. stand for magnetic and electrical fields, respectively, whereas i.p. and o.o.p stand for in plane and out of plane with respect to the TMD layer. We identify optical transitions in red, and transitions between states connected by time-reversal conjugacy in blue.}
\label{TabSI::ML_Selection}
	\begin{tabular}{|c|c|c|c|c|c|}
		\hline
		Symmetry & Physical & Intra-valley & Intra-valley & Inter-valley & Inter-valley\\
		C\textsubscript{3h} & Mechanism & Spin cons. & Spin flip. & Spin cons. & Spin flip. \\ \hline
		$\text{A}'$ & Magn. o.o.p & & & $\text{K}_{\text{VB}} \rightarrow \Gamma_{\text{VB}}$ & \\
		& & & & $\text{K}'_{\text{VB}} \rightarrow \Gamma_{\text{VB}}$ & \\
		& & & & $\text{K(K')}_\text{CB} \rightarrow \text{Q(Q')}_\text{CB}$ & \\ \hline
		
		$\text{A}''$ & El. o.o.p. & & \textcolor{red}{$\text{K}_{\text{VB}}\rightarrow \text{K}_{\text{CB1}}$} & & \textcolor{blue}{$\text{K}_{\text{CB2}} \rightarrow \text{K}'_{\text{CB2}}$} \\
		& & & \textcolor{red}{$\text{K}'_{\text{VB}}\rightarrow \text{K}'_{\text{CB1}}$} & & \\
		& & & & & $(\text{K, K})'_\text{CB} \rightarrow \text{Q(Q')}_\text{CB}$ \\ \hline
		
		$\text{E}'_{+}$ & El. i.p.& \textcolor{red}{$\text{K}_{\text{VB}}\rightarrow \text{K}_{\text{CB2}}$} & & $\text{K}_{\text{CB1}} \rightarrow \text{K}'_{\text{CB2}}$ & \\
		& $\sigma_+$ & & & $\text{K}_{\text{CB2}} \rightarrow \text{K}'_{\text{CB1}}$ & \\
		 & &  & & $\text{K(K')}_\text{CB} \rightarrow \text{Q(Q')}_\text{CB}$ & \\ \hline
		
		$\text{E}''_{+}$ & Magn. i.p. & & $\text{K}_{\text{CB1}} \rightarrow \text{K}_{\text{CB2}}$ & & \textcolor{blue}{$\text{K}'_{\text{VB}}\rightarrow \text{K}_{\text{VB}}$}\\
		& & & $\text{K}'_{\text{CB2}} \rightarrow \text{K}'_{\text{CB1}}$ & & \textcolor{blue}{$\text{K}'_{\text{CB1}} \rightarrow \text{K}_{\text{CB1}}$}\\
		& & & & & $\text{K}'_{\text{VB}}\rightarrow \Gamma_{\text{VB}}$ \\
		& & & & & $\text{K(K')}_\text{CB} \rightarrow \text{Q(Q')}_\text{CB}$ \\ \hline
		
		$\text{E}'_{-}$ & El. i.p. & \textcolor{red}{$\text{K}'_{\text{VB}}\rightarrow \text{K}'_{\text{CB2}}$} & & $\text{K}'_{\text{CB1}} \rightarrow \text{K}_{\text{CB2}}$ & \\
		& $\sigma_-$ & & & $\text{K}'_{\text{CB2}} \rightarrow \text{K}_{\text{CB1}}$ & \\
		& & & & $\text{K(K')}_\text{CB} \rightarrow \text{Q(Q')}_\text{CB}$ & \\ \hline
		
		$\text{E}''_{-}$ & Magn. i.p. & & $\text{K}'_{\text{CB1}} \rightarrow \text{K}'_{\text{CB2}}$ & & \textcolor{blue}{$\text{K}_{\text{VB}}\rightarrow \text{K}'_{\text{VB}}$}\\
		& & & $\text{K}_{\text{CB2}} \rightarrow \text{K}_{\text{CB1}}$  & & \textcolor{blue}{$\text{K}_{\text{CB1}} \rightarrow \text{K}'_{\text{CB1}}$}\\
		& & & & & $\text{K}_{\text{VB}}\rightarrow \Gamma_{\text{VB}}$ \\
		& & & & & $\text{K(K')}_\text{CB} \rightarrow \text{Q(Q')}_\text{CB}$ \\ \hline
	\end{tabular}
\end{table}

\FloatBarrier

\section{Electronic eigenstates and selection rules for bilayer TMDs}
Here, we repeat the steps described previously for the monolayer, now applied to the electronic states in bilayer TMDs. 

\subsection{Character tables}

When we consider the full symmetry of the bilayer stack, the double groups of interest are $\bar{\text{D}}_\text{3d}$ at the $\Gamma$ point of the BZ, and $\bar{\text{D}}_3$ at the K and K' points. At the Q and Q' points, the double group of interest is $\bar{\text{C}}_2$. In the main text, we explain that it is interesting to look at the bilayer as a stack of two monolayers. In this case, the symmetry of the $\Gamma$ point becomes $\bar{\text{D}}_3$, whereas the symmetry of the K and K' points becomes $\bar{\text{C}}_3$. 
Below, we present the character tables for the point groups of interest here. 

\begin{table}[h]
	\begin{tabular}{|l|c c c c c c c c c|}
		\hline
		$\text{D}_\text{3d}$ & $\text{E}$ & $\bar{\text{E}}$ & $2 C_3$ & $2 \bar{C}_3$ & $3 C_2$ & $i$ & $2 S_6$ & $2 \bar{S}_6$ & $3\sigma_d$\\ \hline
		$\Gamma_1$ $(\text{A}_{1g})$ &1 &1 &1 &1 &1 &1 &1 &1 &1\\
		$\Gamma_2$ $(\text{A}_{2g})$ &1 &1 &1 &1 &-1 &1 &1 &1 &-1  \\
		$\Gamma_3$ $(\text{E}_g)$ &2 &2 &-1 &-1 &0 &2 &-1 &-1 &0 \\
		$\Gamma_4$ $(\text{A}_{1u})$ &1 &1 &1 &1 &1 &-1 &-1 &-1 &-1 \\
		$\Gamma_5$ $(\text{A}_{2u})$ &1 &1 &1 &1 &-1 &-1 &-1 &-1 &1 \\
		$\Gamma_6$ $(\text{E}_u)$ &2 &2 &-1 &-1 &0 &-2 &1 &1 &0 \\ \hline
		$\Gamma_7$ $(\bar{\text{E}}_{1/2})$ &2 &-2 &1 &-1 &0 & 0 &$\sqrt{3}$ &$-\sqrt{3}$ &0 \\
		$\Gamma_8$ $(\bar{\text{E}}_{3/2})$ &2 &-2 &-2 &2 &0 &0 &0 &0 &0\\
		$\Gamma_9$ $(\bar{\text{E}}_{5/2})$ &2 &-2 &1 &-1 &0 & 0 &$-\sqrt{3}$ &$\sqrt{3}$ &0 \\ \hline
	\end{tabular}
\end{table}

\begin{table}[h]
	\begin{tabular}{|l|c c c c c c|}
		\hline
		$\text{D}_{3}$ & $\text{E}$ & $\bar{\text{E}}$ & $2 C_3$ & $2 \bar{C}_3$ & $3 C_2$ & $3 \bar{C}_2 $ \\ \hline
		$\text{K}_1$ $(\text{A}_1)$ &1 &1 &1 &1 &1 &1 \\
		$\text{K}_2$ $(\text{A}_2)$ &1 &1 &1 &1 &-1 &-1\\
		$\text{K}_3$ $(\text{E})$ &2 &2 &-1 &-1 &0 &0\\ \hline
		$\text{K}_4$ $(\bar{\text{E}}_{1/2})$ &2 &-2 &1 &-1 &0 &0\\
		$\text{K}_5$ $(\bar{\text{E}}_{-3/2})$ &1 &-1 &-1 &1 &$i$ &$-i$\\
		$\text{K}_6$ $(\bar{\text{E}}_{+3/2})$ &1 &-1 &-1 &1 &$-i$ &$i$\\ \hline
	\end{tabular}
\end{table}

\begin{table}[h]
	\begin{tabular}{|l|c c c c|}
		\hline
		C\textsubscript{2} & $\text{E}$ & $C_2$ & $\bar{\text{E}}$ & $\bar{C}_2$ \\ \hline
		$\text{Q}_1$ $(\text{A})$ &1 &1 &1 &1\\
		$\text{Q}_2$ $(\text{B})$ &1 &-1 &1 &-1\\ \hline
		$\text{Q}_3$ $(\bar{\text{E}}_{-3/2})$ &1 &$i$ &-1 & $-i$\\
		$\text{Q}_4$ $(\bar{\text{E}}_{+3/2})$ &1 &$-i$ &-1 & $i$\\ \hline
	\end{tabular}
\end{table}

\begin{table}[h]
	\begin{tabular}{|l|c c c c c c|}
		\hline
		C\textsubscript{3} & $E$ & $\bar{E}$ & $C_3^+$ & $C_3^-$ & $\bar{C}_3^+$ & $\bar{C}_3^-$ \\ \hline
		$\text{K}_1$ $(\text{A})$ &1 &1 &1 &1 &1 &1 \\
		$\text{K}_2$ $(\text{E}_+)$ &1 & 1 & $\omega$ &$\omega^*$ & $\omega$ &$\omega^*$ \\
		$\text{K}_3$ $(\text{E}_-)$ &1 &1 & $\omega^*$ &$\omega$ & $\omega^*$ &$\omega$ \\ \hline
		$\text{K}_4$ $(\bar{\text{E}}_{-1/2})$ &1 & -1 & $-\omega$ &$-\omega^*$ & $\omega$ &$\omega^*$ \\
		$\text{K}_5$  $(\bar{\text{E}}_{1/2})$ &1 &-1 & $-\omega^*$ &$-\omega$ & $\omega^*$ & $\omega$  \\
		$\text{K}_6$ $(\bar{\text{E}}_{3/2})$ &1 &-1 &-1 &-1 &1 & 1  \\ \hline
	\end{tabular}
\end{table}

\FloatBarrier

\subsection{Band structure and symmetries}

As we did previously for the monolayer, in Tab.~\ref{TabSI::BL_GWV} we present the symmetry group associated with the different points of the BZ (PG column), the atomic orbital character of wavefunctions at these points, and their symmetries (Atomic Orbital and $\Gamma_\text{orb}$ columns respectively), the IR associated with the equivalent representation for wavefunctions at these points ($\Gamma_\text{equiv}$) and, lastly, the full symmtery of the spatial part of the Bloch wavefunctions at these points of the BZ ($\Gamma_\text{spatial}$). 

In tables \ref{TabSI::BL_VB} and \ref{TabSI::BL_CB} we present the symmetry of the Bloch wavefunctions at the high-symmetry points of the BZ including spin, as well as the compatibility relations necessary to obtain the selection rules for scattering between points of the BZ with different symmetries.

\begin{table}[h]
\caption{Spatial character of the wavefunctions at edges of CB and VB at various points of the BZ of bilayer TMDs and their symmetries.}
\label{TabSI::BL_GWV}
	\begin{tabular}{|c|c|c|c|c|c|}
		\hline
		BZ point & PG & Atomic Orbital & $\Gamma_{\text{orb}}$ & $\Gamma_\text{equiv}$ & $\Gamma_\text{spatial} = \Gamma_\text{equiv} \otimes \Gamma_{\text{orb}}$ \\ \hline
		
		$\Gamma$, VB & D\textsubscript{3d} &  d\textsubscript{z\textsuperscript{2}} & $\text{A}_{1g}$ & $\text{A}_{1g}$ & $\text{A}_{1g}$ \\ \hline
		
		K (\text{K}'), VB & D\textsubscript{3} & d\textsubscript{xy, x\textsuperscript{2}-y\textsuperscript{2}} & $\text{E}$ & $\text{E}$ & $\text{A}_1 + \text{A}_2 + E$ \\ 
		
		K (\text{K}'), CB & D\textsubscript{3} & d\textsubscript{z\textsuperscript{2}} & $\text{A}_1$ & $\text{E}$ & $\text{E}$ \\ \hline
		
		Q (Q'), CB & C\textsubscript{2} & d\textsubscript{xy, x\textsuperscript{2}-y\textsuperscript{2}} & $2 \text{A}$ & $\text{A}$ & $2\text{A}$ \\ \hline
	\end{tabular}
\end{table}

\begin{table}[h]
\caption{For valence band wavefunctions, we can obtain the symmetry of the electronic eigenstates including spin by considering the IRs of the double groups. We present also the compatibility relations for IRs of the double groups D\textsubscript{3d}, D\textsubscript{3} and C\textsubscript{2} and C\textsubscript{3}.}
\label{TabSI::BL_VB}
	\begin{tabular}{|c|c|c |c c c|c|}
		\hline
		BZ point & VB wf & $\Gamma_{\text{orb}} \otimes \Gamma_{\text{spin}}$ & $D_{3d}$ & $D_{3}$ & $C_{2}$ & $C_3$ \\ \hline
		$\Gamma$ & $\text{A}_{1g} \otimes \text{E}_{1/2}$ & $\Gamma_7$ & & $\text{K}_4$ & $\text{Q}_3 + \text{Q}_4$ & 2 ($\text{E}_{-1/2} + \text{E}_{1/2}$)\\
		K & $\text{A}_1 \otimes \text{E}_{1/2}$ & $\text{K}_4$ & $\Gamma_7$ & & $(\text{Q}_3+\text{Q}_4)$ & 2 ($\text{E}_{-1/2} + \text{E}_{1/2}$)\\ \hline
	\end{tabular}
\end{table}

\begin{table}[h]
\caption{For conduction band wavefunctions, we can obtain the symmetry of the electronic eigenstates including spin by considering the IRs of the double groups. We present also the compatibility relations for IRs of the double groups D\textsubscript{3d}, D\textsubscript{3} and C\textsubscript{2} and C\textsubscript{3}.}
\label{TabSI::BL_CB}
	\begin{tabular}{|c|c|c|c c c|c|}
		\hline
		BZ point & CB wf & $\Gamma_{\text{orb}} \otimes \Gamma_{\text{spin}}$ & $D_{3d}$ & $D_{3}$ & $C_{2}$ & $C_3$ \\ \hline
		Q & $2A \otimes (\text{Q}_3+\text{Q}_4)$ & $2(\text{Q}_3 + \text{Q}_4)$ & & & $2(\text{Q}_3 + \text{Q}_4)$ & 2 ($\text{E}_{-1/2} + \text{E}_{1/2}$)\\
		K & $E \otimes \text{E}_{1/2}$ & $\text{K}_4 + \text{K}_5 + \text{K}_6$ & $\Gamma_7 + \Gamma_8$ & & $2(\text{Q}_3+\text{Q}_4)$ & 2 ($\text{E}_{-1/2} + \text{E}_{1/2}$) + 2 $\text{E}_{3/2}$ \\ \hline
	\end{tabular}
\end{table}

\begin{figure}
	\centering
	\includegraphics[width=\linewidth]{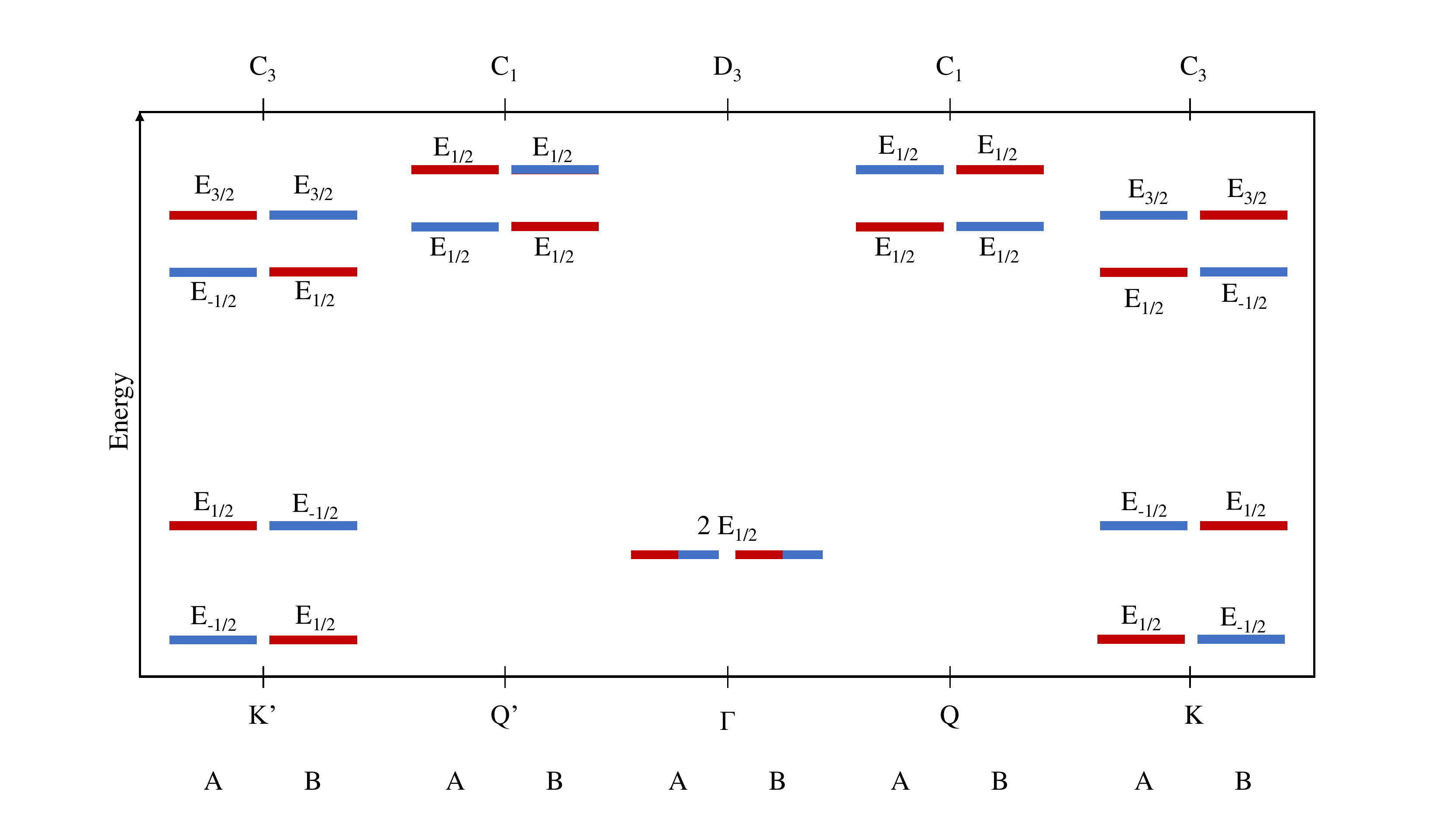}\\
	\caption{\textbf{Energy diagram BL} Blue lines represent spin down states, whereas red lines represent spin up states. Lines of both colors represent degeneracies protected by symmetry. The IRs associated with each state are those of the group of the wave vector at each high symmetry point, indicated at the top. State's localization at each monolayer is indicated by the A and B label at the bottom.}
		\label{fig:Fig2}
\end{figure}

\subsection{Possible scattering processes}

Here, we derive the allowed scattering processes for electrons and holes at the high-symmetry points of the BZ.
We focus specifically on processes leading to inter-layer scattering processes, since the intra-layer processes are analogous to those seen in monolayer TMDs (Sec. \ref{subsec::ML_Selection}). 
Again, we highlight optical transitions in red. 
Since electronic states localized in different layers cannot be each other's time-reversal conjugates, here we observe no transitions that are protected by time-reversal symmetry. 

\begin{table}[h]
	\caption{Inter-layer scattering mechanisms for electrons in K and K' points in bilayer TMD. The abbreviations Magn. and El. stand for magnetic and electrical fields, respectively, whereas i.p. and o.o.p stand for in plane and out of plane with respect to the TMD layers. We identify optical transitions in red.}
	\begin{tabular}{|c|c|c|c|c|c|}
		\hline
		Symmetry & Physical & Intra-valley & Intra-valley & Inter-valley & Inter-valley\\
		C\textsubscript{3} & Mechanism & Spin cons. & Spin flip. & Spin cons. & Spin flip. \\ \hline
		$\text{A}$ & Magn. o.o.p & & $\text{K}_{\text{CB2}} \rightarrow \text{K}_{\text{CB2}}$ & $\text{K}_{\text{VB}}\rightarrow \text{K}'_{\text{VB}}$ & \\
		& El. o.o.p. & & & $\text{K}_{\text{CB1}} \rightarrow \text{K}'_{\text{CB1}}$ & \\
		& & & & $\text{K}_{\text{CB2}} \rightarrow \text{K}'_{\text{CB2}}$ & \\
		& & & & $\text{K}_{\text{VB}}\rightarrow \Gamma_{\text{VB}}$ &\\
		& & & & $\text{K(K')}_\text{CB}\rightarrow \text{Q(Q')}_\text{CB}$ & $\text{K(K')}_\text{CB} \rightarrow \text{Q(Q')}_\text{CB}$ \\ \hline
		
		$\text{E}_+$ & Magn. i.p. & \textcolor{red}{$\text{K}_{\text{VB}}\rightarrow \text{K}_{\text{CB1}}$} & \textcolor{red}{$\text{K}_{\text{VB}}\rightarrow \text{K}_{\text{CB2}}$} & &  \\
		& El. i.p. & & $\text{K}_{\text{VB}}\rightarrow \text{K}_{\text{VB}}$ & & \\
		& & & $\text{K}_{\text{CB1}} \rightarrow \text{K}_{\text{CB1}}$ &  & \\
		& & $\text{K}_{\text{CB1}} \rightarrow \text{K}_{\text{CB2}}$ & & & $\text{K}_{\text{CB1}} \rightarrow \text{K}'_{\text{CB2}}$ \\ 
		& & & & & $\text{K}_{\text{VB}}\rightarrow \Gamma_{\text{VB}}$ \\ \hline
		
		$\text{E}_-$ & Magn. i.p. & \textcolor{red}{$\text{K}'_{\text{VB}}\rightarrow \text{K}'_{\text{CB1}}$} & \textcolor{red}{$\text{K}'_{\text{VB}}\rightarrow \text{K}'_{\text{CB2}}$} & &  \\
		& El. i.p. & & $\text{K}'_{\text{VB}}\rightarrow \text{K}'_{\text{VB}}$ & & \\
		& & & $\text{K}'_{\text{CB1}} \rightarrow \text{K}'_{\text{CB1}}$ &  & \\
		& & $\text{K}'_{\text{CB1}} \rightarrow \text{K}'_{\text{CB2}}$ & & & $\text{K}'_{\text{CB1}} \rightarrow \text{K}_{\text{CB2}}$ \\ 
		& & & & & $\text{K}'_{\text{VB}}\rightarrow \Gamma_{\text{VB}}$ \\ \hline

	\end{tabular}
\end{table}

\newpage

\noindent \textbf{References for Supplementary Information}
\bibliography{myBib}